\documentclass[11pt,twoside]{article}

\usepackage{asp2006}
\usepackage{epsf}
\usepackage{lscape}

\begin{document}
\title{The pattern speeds of NGC 6946}  
\author{Silvia Toonen$^{1}$, Kambiz Fathi$^{2}$, Jes\'{u}s Falc\'{o}n-Barroso$^{1,3}$, John Beckman$^{2}$, and Tim de Zeeuw$^{1}$}   
\affil{1) Leiden Observatory, Leiden University, The Netherlands\newline
2) Instituto de Astrof\'{i}sica de Canarias, La Laguna, Spain\newline
3) ESTEC, Noordwijk, The Netherlands}   

\begin{abstract} 
We study the kinematics of the barred spiral galaxy NGC 6946 by investigating the velocity field from H$\alpha$ Fabry-Perot observations, determined the pattern speed of the bar by using the Tremaine-Weinberg method, and find a main pattern speed of 21.7 (+4.0,-0.8) km/s/kpc. Our data clearly suggest the presence of an additional pattern with a pattern speed more than twice that of the large pattern in this galaxy. We use the epicycle approximation to deduce the location of the resonance radii and subsequently determine the pattern speed between the radii, and find that inside the Inner Inner Lindblad Resonance radius, a bar-like system has evolved. 
\end{abstract}

\section{Introduction}
The study of the ISM and its kinematics enables us to address one of the more interesting questions in galactic evolution: by what mechanism or mechanisms is the optical appearance of a galaxy related to the evolution of its spiral or barred structure in response to the underlying dynamics of its stellar and gaseous components, and what is the role of such evolution on star formation in the host galaxy? At a distance of 5.5 Mpc, NGC 6946 provides us with a good opportunity to study and characterize the properties of interstellar gas and dust in detail. The observations were taken with the Fabry-Perot interferometer FaNTOmM (Hernandez, Gach, Carignan \& Boulesteix 2003) at the 1.6-m telescope at the Observatoire du Mont M$\mathrm{\acute{e}}$gantic, Qu$\mathrm{\acute{e}}$bec, Canada. The resulting data cube of 40 planes separated by 0.21 $\mathrm{\AA}$ was used to derive flux, line-of-sight velocity and velocity dispersion for each position. 
The radial velocity map shows the presence of a small, asymmetrically extended, inner bar-like system embedded in a large-scale rotating structure. 

\begin{figure}[!htb]
\begin{center}
\begin{tabular}{c c}
\scalebox{0.35}{\includegraphics{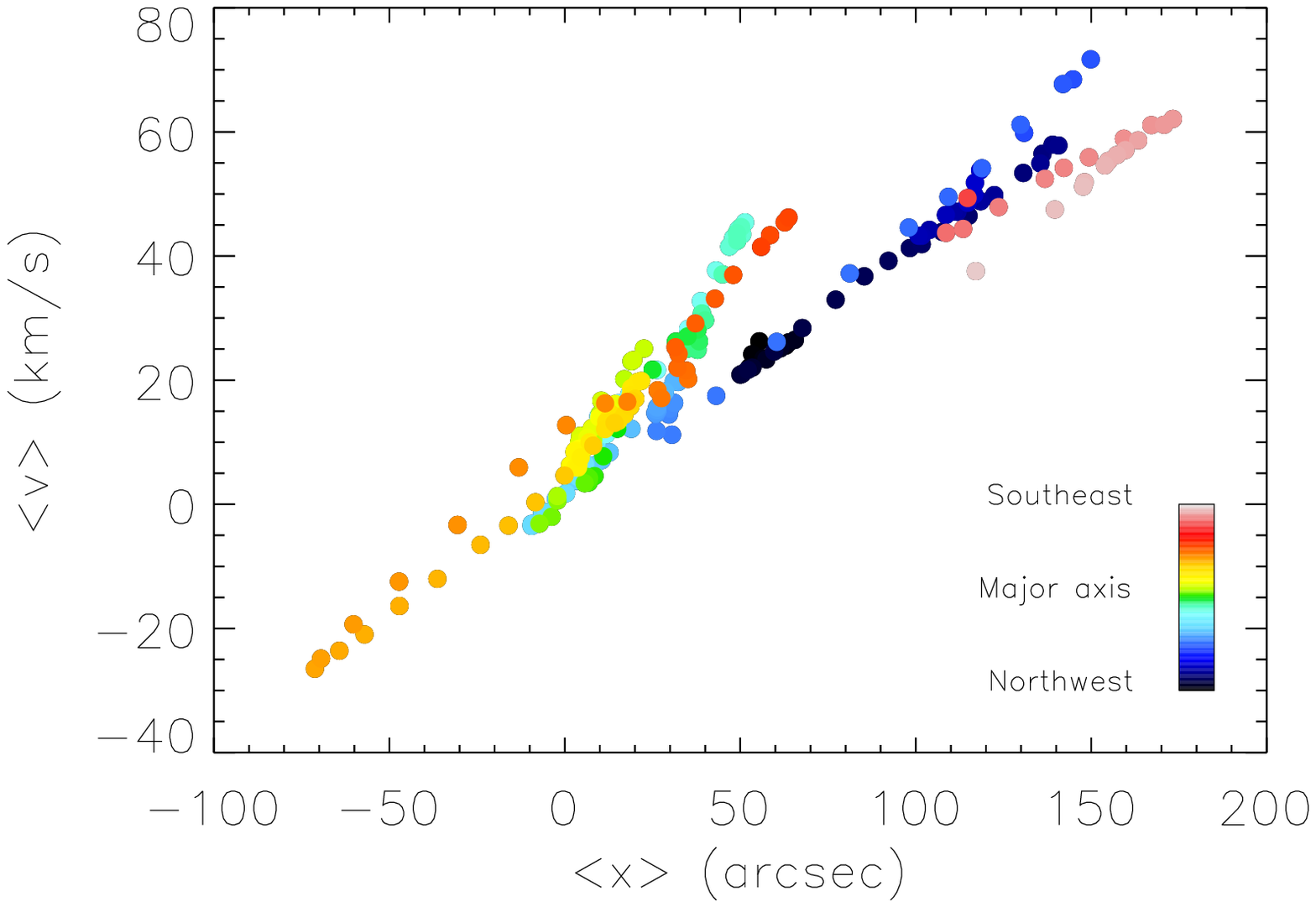}} &
\scalebox{0.35}{\includegraphics{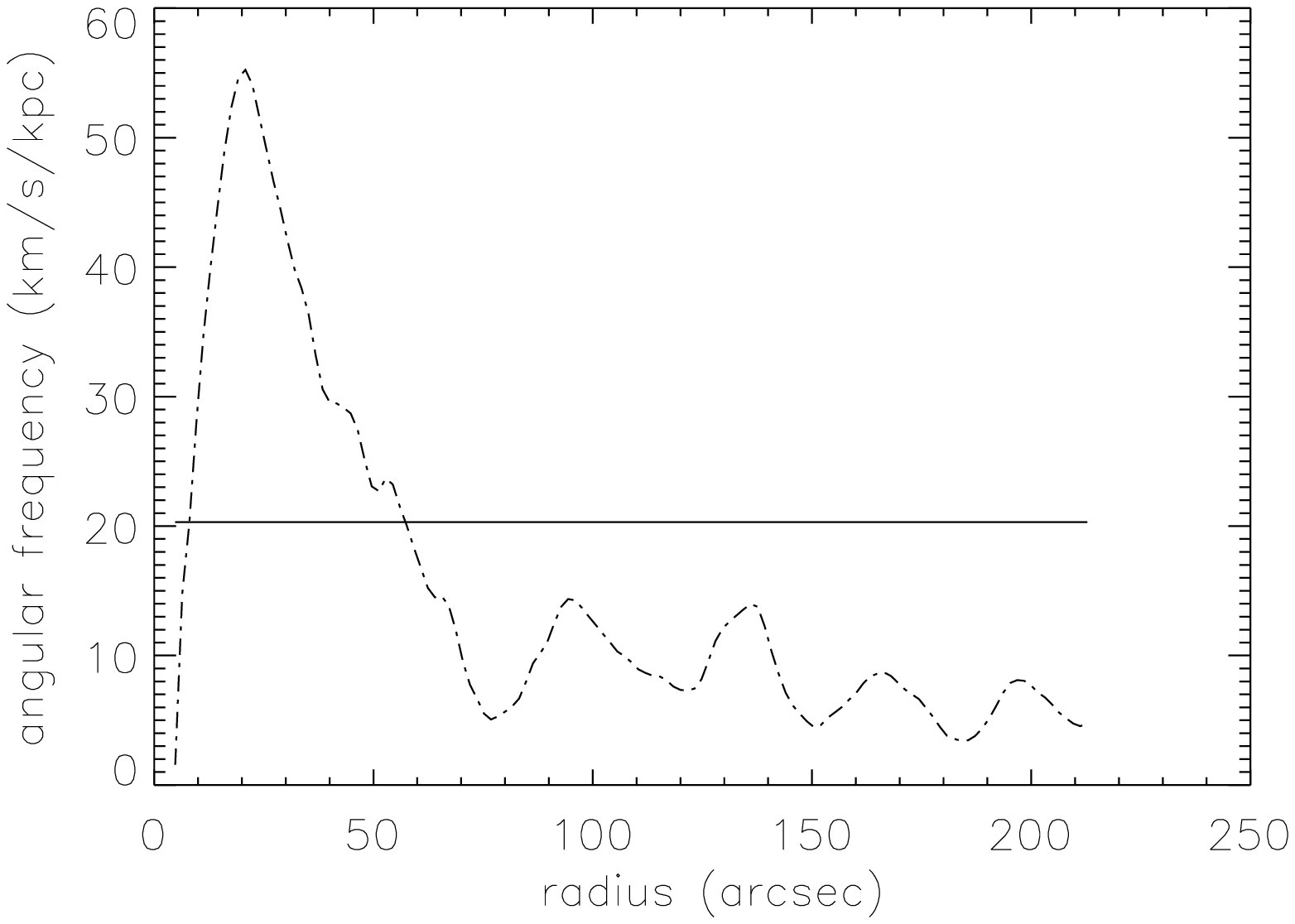}} \\
\small \emph{(a)} & \small \emph{(b)}\\
\end{tabular}
\caption{\emph{(a) Pattern speed determination with the Tremaine \& Weinberg method. The pattern speed is given by the ratio of $\langle$v$\rangle$ and $\langle$x$\rangle$. Measurements through the centre are distributed along the line of higher slope, where the measurements solely through the exterior of the NGC 6946 are accumulated along the line of lower slope, indicating two distinct pattern speeds. (b) The $(\Omega-\frac{\kappa}{2})$-curve, with $\Omega$ the angular frequency and the $\kappa$ the epicyclic frequency, indicating two ILRs. }}
\label{fig:plaatjes}
\end{center}
\end{figure}

\section{Two pattern speeds}
The Tremaine \& Weinberg method (1984) makes it possible to measure the pattern speed of a barred disk galaxy without adopting a particular dynamical model. It is based on the ratio of measurements of the distribution of intensityweighted velocity and position of a component that reacts to the density wave. We use this method with the refinement of Merrifield \& Kuijken (1995). The distribution of the measurements along two lines in Fig.\ 1a suggests not one, but \textit{two} distinct pattern speeds for measurements including and excluding the central regions. This observational result suggests the presence of an additional, inner pattern that rotates at a different pattern speed roughly inside the OILR. The outer pattern rotates at 21.7 (+0.5,-0.5) km/s/kpc and the inner pattern rotates at 47.0 (+1.4,-1.4) km/s/kpc.

\section{Resonance radii}
We use the epicycle approximation to deduce the location of the Lindblad Resonance radii (ILR) and the Corotation Radius. Fig.\ 1b shows the ($\Omega-\frac{\kappa}{2}$)-curve. This indicates two ILRs; an IILR and OILR. As described by Maciejewski (2004), the curve can be interpreted as representing a galaxy without a central mass concentration or density cusp. Fig. 1b also indicates that a bar-like system with solid-body rotation has evolved inside the IILR as confirmed by the steeply rising rotation curve in this galaxy.\newline \newline
In a forthcoming paper, we will present an extended analysis of the pattern speeds and a diagnosis of the resonance radii, within the framework of epicycle approximation, in NGC 6946.


\end{document}